\documentclass[letter]{aa} 

\usepackage{graphicx}
\usepackage{hyperref}
\usepackage{multirow}

\newcommand{\lum}{erg~s\ensuremath{^{-1}}}

\newcommand{\flux}{erg~s$^{-1}$~cm$^{-2}$}

\newcommand{\msun}{\ensuremath{M_{\odot}}}

\usepackage{txfonts}

\begin{document}

\title{X-ray view of a merging supermassive black hole binary candidate SDSSJ1430+2303: Results from the first $\sim$200 days of observations} 

\titlerunning{X-rays of SDSSJ1430+2303}

%

\author{Liming Dou\inst{1} 
          \and
          Ning Jiang\inst{2,3}
          \and
         Tinggui Wang\inst{2,3}
            \and
          Xinwen Shu\inst{4}
           \and 
           Huan Yang\inst{5,6}
           \and
           Zhen Pan\inst{5}
           \and
           Jiazheng Zhu\inst{2,3}
           \and
          Tao An\inst{7,8} 
           \and
           Zhen-Ya Zheng\inst{7}    
          \and
          Yanli Ai\inst{9}} 

\institute{Department of Astronomy, Guangzhou University, Guangzhou 510006, China \\
              \email {doulm@gzhu.edu.cn}
          \and
          CAS Key laboratory for Research in Galaxies and Cosmology, Department of Astronomy, University of Science and Technology of China, Hefei, 230026, China\\ 
            \email{jnac@ustc.edu.cn,twang@ustc.edu.cn}
            \and
            School of Astronomy and Space Sciences, University of Science and Technology of China, Hefei, 230026, China
            \and
            Department of Physics, Anhui Normal University, Wuhu, Anhui, 241002, China
            \and 
            Perimeter Institute for Theoretical Physics, Waterloo, ON N2L2Y5, Canada
            \and
             University of Guelph, Department of Physics, Guelph, ON N1G2W1, Canada        
            \and
            Shanghai Astronomical Observatory, Chinese Academy of Sciences, Nandan Road 80, Shanghai 200030, China       
            \and
            SKA Regional Centre Joint Lab, Peng Cheng Laboratory, Shenzhen, 518066, China
            \and
            College of Engineering Physics, Shenzhen Technology University, Shenzhen 518118, PR China
            }

   \date{Received 7/7/2022; accepted }


\abstract
{Recently we discovered an unprecedented supermassive black hole binary (SMBHB) candidate in the nearby Seyfert galaxy SDSS J1430+2303, which is predicted to merge within three years. X-ray spectroscopy may bring unique kinematic evidence for the last inspiraling stage, when the binary is too close to allow each of them to hold an individual broad line region. }
{We try to confirm the unique SMBHB merger event and understand the associated high-energy processes from a comprehensive X-ray view. }
{We observed SDSS J1430+2303 with {\it XMM-Newton}, {\it NuSTAR}, {\it Chandra,} and {\it Swift} spanning the first $\sim200$ days since its discovery.}
{X-ray variability, up to a factor of 7, has been detected on a timescale of a few days. The broadband spectrum from 0.2--70~keV can be well fitted with a model consisting of a power law and a relativistic reflection covered by a warm absorber. The properties of the warm absorber changed dramatically, for example, with a decrease in the line-of-sight velocity from $\sim$0.2c to $\sim$0.02c, between the two {\it XMM-Newton} observations separated by only 19 days,  which can be naturally understood in the context of the SMBHB; although, the clumpy wind scenario cannot be completely excluded. Broad Fe K$\alpha$ emission has been robustly detected, though its velocity shift or profile change is not yet measurable. Further longer X-ray observations are highly encouraged to detect the expected orbital motion of the binary. }
{}


   \maketitle
%

\section{Introduction}

Supermassive black hole binaries (SMBHBs) are an inevitable and fascinating byproduct of galaxy mergers in the hierarchical Universe~\citep{Begelman1980} since most massive galaxies are expected to contain at least a central SMBH with a mass of  $10^{6}-10^{10}~\msun$~\citep{KH13}. When two galaxies merge, the two SMBHs initially residing in the center of each galaxy sink to the common center of the merged system via dynamic friction. SMBHBs at this stage may be identified as dual active galactic nuclei (AGNs) with separations ranging from several parsecs to several kiloparsecs (e.g., \citealt{Zhou2004,Liu2010,Comerford2015}). As the two SMBHs continue to shrink their separation by ejecting stars in a "loss cone," they gradually become a gravitationally bound system on subparsec scales. Their orbital angular momentum is further lost due to some other mechanisms (e.g., \citealt{Ivanov1999,Hayasaki2009}) until gravitational wave (GW) radiation takes over. The final coalescence of SMBHBs is thought to be the loudest GW sirens~\citep{Thorne1976,Haehnelt1994,Jaffe2003}, which are the primary targets of ongoing and upcoming GW experiments, such as of the Pulsar Timing Arrays (PTAs) and  the Laser Interferometer Space Antenna (LISA). 

In spite of the attractive prospects, close SMBHBs with separations below {\rm a parsec} scale are however extremely challenging to unveil as they are far beyond the resolution limit of the current generation of telescopes, except for the very few nearby galaxies for which long baseline radio interferometry observations are possible~\citep{Gallimore2004}. The search of subparsec binaries thus must rely on some indirect measurements, such as the shift of broad emission lines due to its orbit motion (\citealt{Eracleous2012,Shen2013,Runnoe2017}), the presence of double-peaked emission profiles from the coexisting broad-line regions associated with individual accreting SMBHs (e.g., \citealt{Boroson2009}), or a drop in the UV continuum due to a gap opened by the secondary BH migrating within the circumbinary disk~\citep{Gultekin2012,Yan2015}. Another promising method is the periodic variations of AGNs, which reveals the possible presence of the SMBHB system in the blazar OJ~287~\citep{Sillanpaa1988,Lehto1996}, and it has become particularly popular recently in virtue of modern time-domain surveys. The periodicity can also be understood as either accretion rate fluctuations~\citep{MacFadyen2008,Noble2012,Graham2015} or relativistic Doppler modulation~\citep{D'Orazio2015}, on a timescale comparable to the orbit period. Although a mounting number of candidates have been selected (e.g., \citealt{Graham2015b,Liu2016,Charisi2016,Zheng2016,Chen2020}), none of them are really approaching its final coalescence such that the observed periods stay constant in time.  

Recently, we have discovered a possible candidate SMBHB close to the merger stage in a nearby galaxy SDSS~J143016.05+230344.4 (hereafter SDSSJ1430+2303) at $z=0.08105$~\citep{Jiang2022}. SDSSJ1430+2303 was initially noticed during our systematic search for mid-infrared (MIR) outbursts in nearby galaxies (MIRONG, \citealt{Jiang2021}). A careful check of its optical light curves from the Zwicky Transient Facility (ZTF) then revealed unique chirping flares, with a reduced period from about one year to only three months by the end of 2021. The flares can be ideally interpreted as emissions from plasma balls that are kicked out from the primary SMBH accretion disk by an inspiraling secondary SMBH during disk crossings. We have developed a trajectory model to explain the period evolution and predicted that the binary would merge within three years, making multiwavelength follow-up observations rather pressing and exciting. Since its discovery, we have triggered intensive follow-up observations in multi-wavelength regimes (e.g., \citealt{An2022a,An2022b}). Among them, X-ray observations are essential for confirming the SMBHB scenario and understanding the associated high-energy physical processes happening during the last insprialing stage. In particular,  the binary is so compact that individual SMBHs are not allowed to possess their own broad line regions, but they can still have their own accretion disks. The orbital motion of SMBHs causes a unique velocity shift of the emission lines in the reflected light of the accretion disk or of the absorption lines from the accretion disk wind. If such signatures are detected, they provide independent kinematic evidence for the existence of the SMBHB. In this Letter, we report the X-ray results of SDSSJ1430+2303 observed with {\it Swift}, {\it XMM-Newton}, {\it Chandra,} and {\it NuSTAR} during the first $\sim$200 day follow-ups since its discovery. 

\section{Observations and data reduction}

\subsection{\it Swift}

In addition to the intriguing optical light curves, we also note four epochs of observations from the X-Ray Telescope (XRT) on the Neil Gehrels Swift observatory ({\it Swift} for short), including one targeted observation (ToO\,ID:13234, PI:L{\'o}pez Navas) on 2018 November 24 and another three occasional visits in 2019 and 2020. The four snapshots give a count rate of $0.05-0.1~{\rm cts\,s^{-1}}$, corresponding to an X-ray luminosity of $(0.5-1)\times10^{44}$~\lum\ in 0.3--10 keV, assuming a fixed average photon index ($\Gamma=1.66$). The high X-ray count rate allowed us to monitor its X-ray luminosity with reasonable time.

We note that when we discovered the chirping flares of SDSSJ1430+2303 in 2021 September, the target was however not visible by a ground telescope in night. We immediately triggered the X-ray monitoring program with {\it Swift} after the period of Sun constraint, when SDSSJ1430+2303 still remained unobservable for ground-based telescopes. As an exploratory observing experiment, we first initiated a one-month-long monitoring from 2021 November 23 to December 21 with a three-day cadence, and each observation has an exposure time of $\sim$1 ks (PI: Jiang, ToO\,ID:16602). With the {\it Swift} observations, we found a large amplitude of X-ray variability ($0.01-0.1~{\rm cts\,s^{-1}}$) with a seemingly even shortened period, which has decreased to only one month. Thus we chose to change the observing strategy to a daily cadence in the subsequent request after December 25. However, {\it Swift} science instruments went into safe mode because of an unexpected reaction wheel failure\footnote{https://swift.gsfc.nasa.gov/news/2022/safe\_mode.html} between 2022 January 18 and February 18. We restarted the daily monitoring when {\it Swift} returned to science operations, which was then followed with a two-day cadence up to now. 

We reprocessed the {\it Swift}/XRT observations following the standard data reduction using tools in HEASOFT (v.6.30) with the latest calibrations. The event files have been rebuilt by the task {\tt xrtpipeline}, with only the observations operated in "photon counting" mode being used. We extracted source photons using a circular region with a radius of $47.2''$ centered on the target and background photons from an annulus region free of source emission. The net count rates in the 0.3--10 keV band for each observation were then calculated and the corresponding count rate light curve was constructed. 

\subsection{\it XMM-Newton}

During the {\it Swift} daily monitoring, we also triggered two epochs of {\it XMM-Newton} DDT observations on 2021 December 31 and 2022 January 19, respectively (obsid: 0893810201 \& 0893810401).  We reprocessed the {\it XMM-Newton}/EPIC data with the Science Analysis Software (SAS, v.20) and the latest calibration files. Only the data from the pn instrument of EPIC were used for our analysis in view of its higher sensitivity than MOS instruments. We created the events files with the tool of {\tt epchain}. After removing the "bad" pixels, we created the high flaring particle background time intervals with a threshold rate of $>0.6~{\rm cts\,s^{-1}}$  with single events ("PATTERN==0") in the 10--12 keV band. This resulted in a net exposure time of 36.06 and 40.54 ks for the two observations, respectively. We used only the single events for the science analysis. We extracted the source spectra from a circular region with a radius of $40"$ centered on its optical position, and the background spectra from a nearby source-free circular region with the same radius. 

\subsection{\it NuSTAR}

A {\it NuSTAR} ToO observation was performed from 2022 February 3-6 with a total exposure time of 102.9 ks during an observation span of 212 ks. We reprocessed the {\it NuSTAR} data using the latest calibration files with the tools in HEASOFT. We used the task {\tt nupipeline} to filter the event lists with the options of "saacalc=3" and "saamode=strict," and used the task {\tt nuproducts} to extract spectra and response files for the focal plane modules (FPMA and FPMB). The source spectrum was  extracted from a circular region with a radius of $40"$ centered on its optical position, and the background spectrum was extracted from a nearby source-free circular region with a larger radius. This resulted in a net 3--79 keV band count rate of 0.062$\pm0.001~{\rm cts\,s^{-1}}$  for FPMA within an efficient exposure of 81.3 ks, and 0.059$\pm0.001~{\rm cts\,s^{-1}}$ for FPMB within an efficient exposure of 80.8 ks. We found no obvious large-amplitude variability during the observation. We then merged the two spectra from FPMA and FPMB together with the tool {\tt addspec} of HEASOFT. 

\subsection{\it Chandra}

We requested 15 {\it Chandra} DDT observations (PI: Jiang N.) which were performed from 2022 February 21 to March 16 with an average exposure time of 4 ks. We reprocessed the {\it Chandra} data with CIAO (v.4.14) and the latest calibration files (v.4.9.6). The level 2 event files were recreated by the script of {\tt chandra\_repro}. The spectra of source and background, as well as the response files, were produced by the script of {\tt specextract}. The source photons were extracted from circular regions with a radius of $7"$-$25"$ depending on the offset between the position of the source and the observation viewing center. Larger source extraction regions were selected because some {\it Chandra} observations were performed off-axis. The background was extracted from larger annulus source-free regions with the same source centers. This resulted in count rates of $0.12-0.23~{\rm cts\,s^{-1}}$ in the 0.5-8 keV. We found no obvious variability within each observation. Finally, we merged all the {\it Chandra} observations into one stacked spectrum, which resulted in an average count rate of $0.16~{\rm cts\,s^{-1}}$ with a total exposure time of 57.4 ks. 

\subsection{\it NICER}

We note that SDSSJ1430+2303 has also been continuously monitored since 2022 January 20 by {\it NICER} (\citealt{Pasham2022}), we will also put them together for a light curve comprehensive analysis. The level 2 {\it NICER}  event files were first reproduced following the standard processing by the task {\tt nicerl2}. After eliminating the high energy background time intervals, the spectrum and light curve of each observation were extracted by the tool of {\tt xselect}.  We used the 3C50 model to estimate the NICER background by using the task {\tt nibackgen3C50.pl}. The background-subtracted count rate of each observation was in the range of 1.1--3.3 ~${\rm cts\,s^{-1}}$ in 0.3--4 keV band. 

 \section{X-ray light curves}

\begin{figure}[htb]
\centering
\begin{minipage}{0.45\textwidth}
\centering
\includegraphics[angle=0,width=1.0\textwidth]{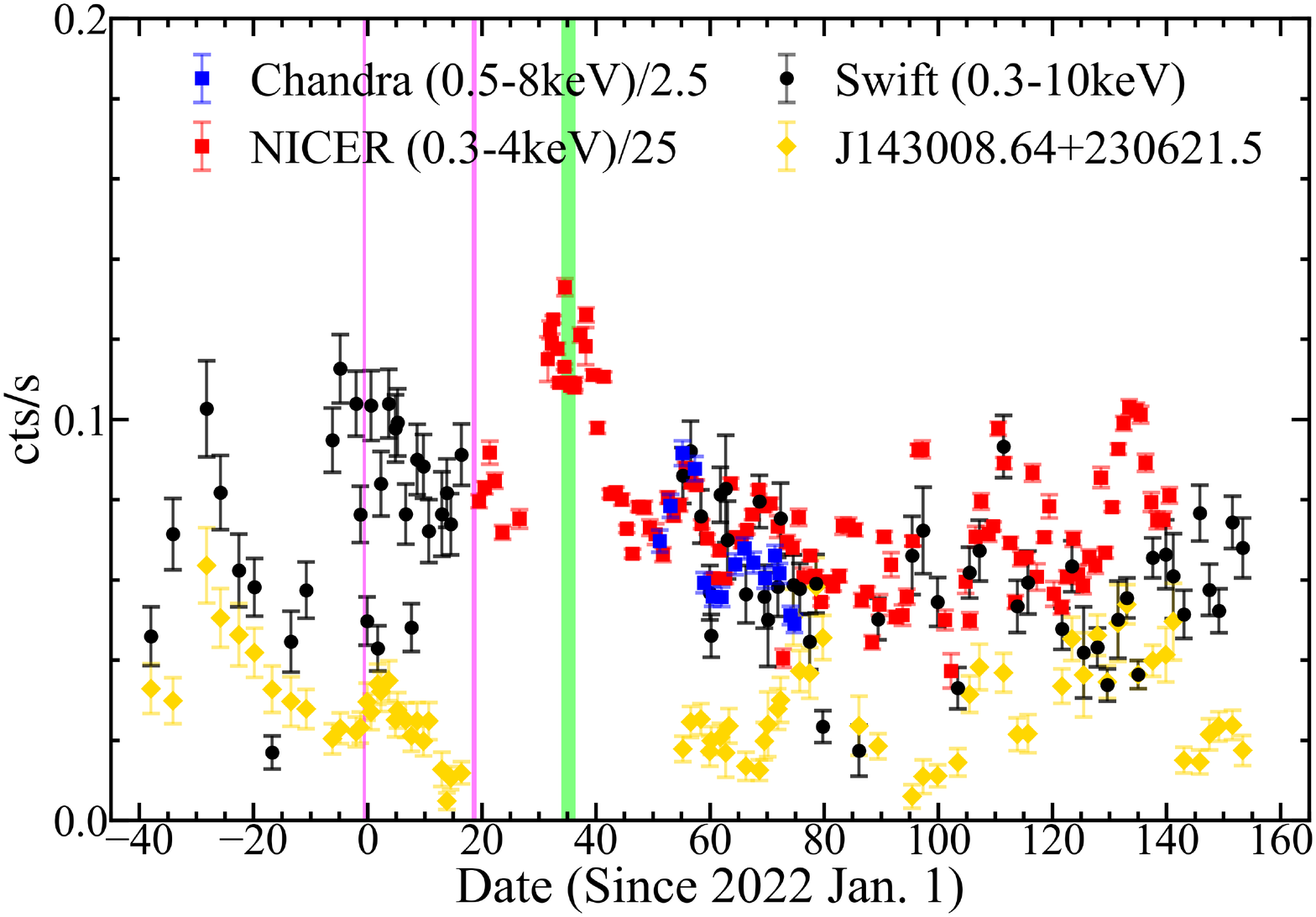}
\end{minipage}
\caption{Long-term X-ray light curves of SDSSJ1430+2303 since the {\it Swift} monitoring campaign. The count rates from {\it Swift,} {\it Chandra,} and {\it NICER} are shown with black dots and blue and red squares, respectively, which have been normalized in the overlapped time range. The {\it Swift} light curve of the nearby quasar, which has contaminated the {\it NICER} measurement, is also overplotted with yellow diamonds. We have also denoted the {\it XMM-Newton} and {\it NuSTAR} observing period with magenta and green vertical shadow regions.\label{xlc}}
\end{figure}

In Figure~\ref{xlc}, we show the X-ray light curves of {\it Swift}/XRT observations in 0.3--10 keV, {\it Chandra}/ACIS-S observations in 0.5--10 keV, and {\it NICER} observations in 0.3--4 keV. In order to compare the variability between different observations, we normalized the count rates of the light curves from {\it Chandra} and {\it NICER} observations to the averaged count rate obtained from the {\it Swift} observations, that is to say from the quasi-simultaneous observations from 2022 February 21 to March 16. 

Although {\it NICER} yielded a much higher count rate, and thus signal-to-noise ratio (S/N), than {\it Swift}/XRT due to its larger effective area, the lack of imaging ability yet a large field of view of 28 {\rm arcmin$^2$}~\citep[]{Arzoumanian2014} makes it difficult to mitigate the contamination of the nearby quasar SDSS J143008.64+230621.5 (SDSSJ1430+2306 hereafter), which also falls into the field view of the target (about 3.1~{\rm arcmin} offset). The quasar has a count rate about 1-200$\%$ of our target and is highly variable in the X-ray band as is illustrated in the {\it Swift}/XRT light curve in Figure~\ref{xlc}.  Thus, we caution that special attention should be paid when analyzing the {\it NICER} light curve or spectrum independently. 

We also generated the background-subtracted light curves for the two {\it XMM-Newton} observations after correcting for the detection efficiency effect with the tool {\tt epiclccorr}. As shown in Figure~\ref{xlc_o1o2}, we have checked the light curves in the 0.2-0.5, 0.5-1, 1-2, and 2-10 keV bands individually, and found no obvious variability on the hour timescale. The average net count rates in the  0.2-10 keV band are both around $1.2~{\rm cts\,s^{-1}}$, though there was a slight rising trend from 1.1 to 1.3~${\rm cts\,s^{-1}}$ during the first visit.  Moreover, the power spectrum analysis did not detect any quasi-periodic oscillation signals with a high confidence. 
 
\section{X-ray spectral analysis}\label{xspect}

 We checked the source and background spectra of the {\it XMM-Newton}, stacked {\it Chandra}, and {\it NuSTAR}  observations, and found that all the source spectra have enough S/N, except for the {\it NuSTAR} spectrum at energies above 35 \,keV, which is dominated by the background. We then regrouped the spectra to ensure there were at least 25 net counts per energy bin for using the $\chi^2$ statistics in the spectral fittings with XSPEC. During the fittings, a neutral absorption column density was included and fixed to the Galactic value ("phabs" in XSPEC, {\rm N$_{H}=2.28\times10^{20} \rm cm^{-2}$}, \citealt{HI4PI2016}), and the uncertainties are given at a 90\% confidence level for one parameter ($\Delta\chi^2=2.706$), unless stated otherwise. We assumed a cosmology with $H_{0} =70$ km~s$^{-1}$~Mpc$^{-1}$, $\Omega_{m} = 0.3$, and $\Omega_{\Lambda} = 0.7$ in this work.

We first applied a simple Galactic absorbed power-law model to fit, jointly, the two {\it XMM-Newton}/pn (marked as xmmobs1 and xmmobs2) in 2-5 and 8-10 keV,  {\it Chandra} stacked spectra in 2-5 keV, and the {\it NuSTAR} merged spectrum in 3-5 and 8-10 keV, while the power-law normalization of each spectrum was set as a free parameter. The fitting bands were chosen so as to mitigate a potential impact of disk reflection or warm absorption. We obtained photon indices of $\Gamma$=1.65$\pm$0.07, 1.61$\pm$0.07, 1.65$\pm$0.10, and 1.71$\pm$0.07 for xmmobs1, xmmobs2, {\it Chandra,} and {\it NuSTAR} spectra, respectively, with a total $\chi^2$=373.6 for 347 degree of freedom (d.o.f). The indices are consistent with each other within uncertainties in spite of a tentative softer-when-brighter trend. If bounding their photon indices together, we obtained an index of 1.66$\pm$0.04, which only increased the total ${\Delta}\chi^{2}$=2.6 for the decrease of three d.o.f. The results from the joint fitting were adopted by us, giving observed $2-10$~keV fluxes of 2.85$\pm$0.05$\times$10$^{-12}$,  2.62$\pm$0.05$\times$10$^{-12}$, 2.51$\pm$0.05$\times$10$^{-12}$, and 3.72$\pm$0.04$\times$10$^{-12}$~\flux\ for xmmobs1, xmmobs2, {\it Chandra,} and {\it NuSTAR} observation, respectively. Albeit with a large amplitude in variability, the two {\it XMM-Newton} epochs were unfortunately observed at almost identical flux levels while the {\it NuSTAR} observation was performed at a stage with a flux 30-40$\%$ higher. These results are consistent with a constant spectral index for observations.  

\begin{figure}[htb]
\centering
\begin{minipage}{0.45\textwidth}
\centering{\includegraphics[angle=0,width=1.0\textwidth]{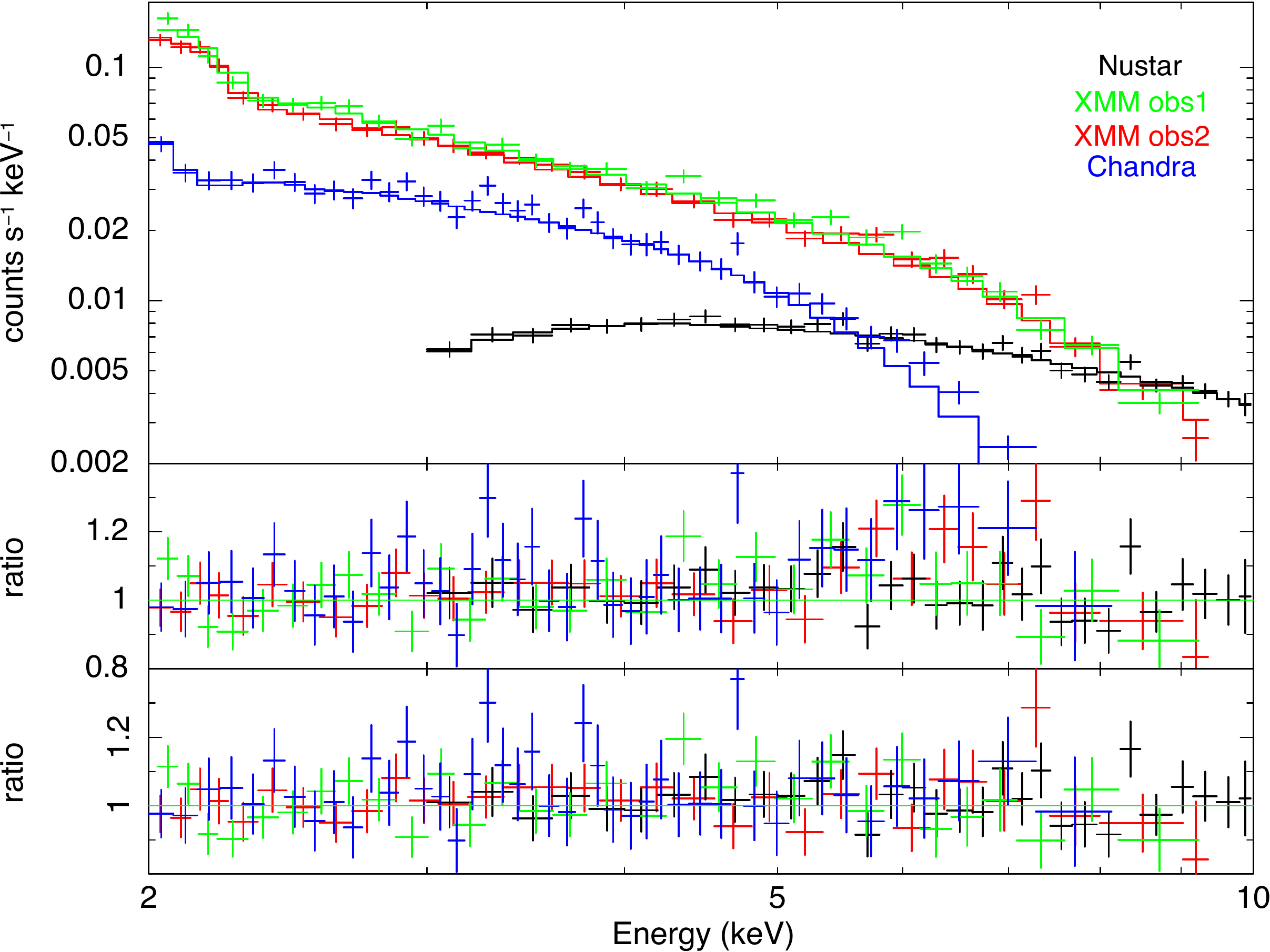}}
\end{minipage}
\caption{Two {\it XMM-Newton}/pn, {\it Chandra} stacked, and {\it NuSTAR} folded spectra in 2-10\,keV with a simple power-law model, and the ratio of the data and model. Upper panel: The model was fitted for the 2-5 and 8-10\,keV band spectra and extrapolated to 2-10\,keV.
Middle panel: The ratio of the data and model with the simple power-law model, and obvious excess in the 5-8 keV band of the two {\it XMM-Newton}/pn and {\it Chandra} observations.
Bottom panel: The ratio of the data and model with the simple power-law plus a Gaussion line model, which might be a relativistic broadening Fe K emission.
\label{fig:excessfe}
}
\end{figure}

However, as shown in the residual spectrum (see the middle panel of Figure\,\ref{fig:excessfe}), obvious excesses in the 5-8 keV band are visible in the two {\it XMM-Newton}/pn and {\it Chandra} observations after the power-law model was interpolated to the above region. We then added a Gaussian line to the model, which yielded a total $\chi^{2}$=560.3 for 538 d.o.f, with ${\Delta}\chi^{2}$=25.0 for the addition of six d.o.f. The addition of a Gaussian line model improved the fits significantly, with a confidence level of 99.96\% as determined through 10,000 
Monte Carlo simulations using the XSPEC script {\tt simftest} \citep[see also][]{Protassov2002}. The fitted Gaussian line is peaked at $E=6.55^{+0.33}_{-1.03}$ keV with a width of $\sigma=0.69^{+1.54}_{-0.28}$ keV. If the line center and width are fixed to the best value, the line detection confidence is 99.1\%, 99.5\%, and 99.3\% for the xmmobs1, xmmobs2, and {\it Chandra} observation, respectively. However, the detection confidence for the independent {\it NuSTAR} spectral fitting is much lower, that is to say 65\%, indicating that the line is weak or not detected. The equivalent width of the Gaussian broad line is 0.22, 0.30, and 0.37~keV for the xmmobs1, xmmobs2, and {\it Chandra} observation, respectively, which is tentatively anticorrelated with the continuum flux, while the parameters of the line normalization are consistent with each other, if considering the uncertainties.  Such a broad Gaussian line component can be naturally treated as a relativistic broadened Fe K$\alpha$ emission line around 6.4~keV \citep[e.g.,][]{Nandra2007}. 

\begin{figure}[htb]
\centering
\begin{minipage}{0.45\textwidth}
\centering{\includegraphics[angle=0,width=1.0\textwidth]{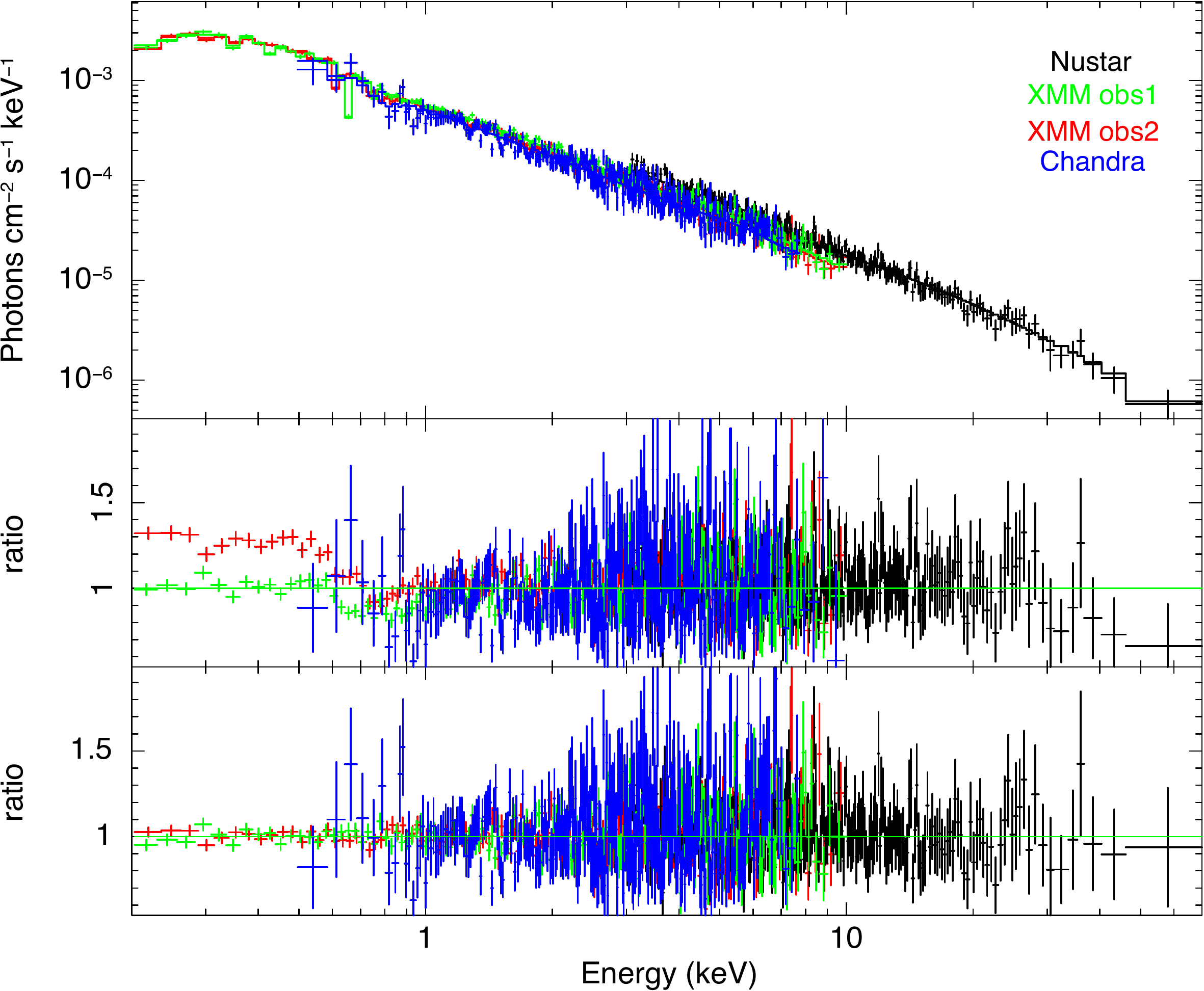}}
\end{minipage}
\caption{ {\it XMM-Newton}, {\it Chandra,} and {\it NuSTAR} spectra with a warm absorbed relativistic reflection model. Upper panel: Unfolded spectra with the best fitted warm absorbed relativistic reflection disk model. Middle panel: Ratio of the data and model with a simple power law plus a Guassion line for the 2-10\,keV band spectra and extrapolated to 0.2-10\,keV of {\it XMM-Newton} and {\it Chandra} spectra and the 3-70\,keV of the {\it NuSTAR} spectrum. Bottom panel: Ratio of the data and model with the best fitted warm absorbed relativistic reflection disk model.
\label{fig:warmabsp}}
\end{figure}

When applying the model to the broadband 3-70 keV spectrum of the {\it NuSTAR} observation and the 0.2-10 keV band spectra of the two {\it XMM-Newton} observations, we found a high-energy cut-off feature above 30 keV in the residual spectrum and apparent deficit features around 0.2-1 keV, as shown in the middle panel of Figure\,\ref{fig:warmabsp}. We then tried to perform joint fits to the broadband spectra of {\it XMM-Newton} in 0.2-10 keV band, the {\it Chandra} stacked spectrum in 0.5-8 keV, and {\it NuSTAR} spectra in the 3-70 keV band, with a relativistic reflection disk model modified by a warm absorber. For the warm absorber, we ran {\tt xstar2xspec} to create a table grid model~\citep["warmabs" in XSPEC,][]{Kallman2001} by assuming a density of n\,=\,10$^8$\,cm$^{-3}$, photon index of $\Gamma$\,=\,1.7, and $v_{turb}$\,=\,3000\,km\,s$^{-1}$. We fixed the intrinsic cosmological redshift of the warm absorber at 0.08105, that is the one measured from the optical spectrum \citep{Jiang2022}. As the {\it NuSTAR} spectrum does not extend to the lower energies (0.2-3 keV) and since the {\it Chandra} spectrum has a low S/N below 1.5 keV, we tied the absorber parameters, that is to say the hydrogen column density ($n_{\rm H}$) and ionization parameter of $log\,[\xi_{w}/ {\rm (erg\,cm\,s^{-1})}]$, in the {\it NuSTAR} and {\it Chandra} spectral fitting to the ones in xmmobs2. On the other hand, for the relativistic reflection \citep["relxill" in XSPEC,][]{Garca2014}, we tied the model parameters to those used to fit the {\it NuSTAR} spectrum, as the {\it XMM-Newton} and {\it Chandra} observations lack data in the 10-79 keV band. There are still too many parameters to constrain, given the degeneracies, so we fixed the inner radius of $R_{\rm in}$ at the marginally stable orbital radius for BH spin {\it a}=0.5, the outer radius of $R_{\rm out}$ at 400$Rg$, and the power-law emissivity at {\it q}=3 for the reflection model. The fitting result is acceptable ($\chi^2/d.o.f$=903.5/847). When replacing the reflection component with a power-law component or removing the warm absorber, the fitting became much worse ($\chi^2/d.o.f$=1074.7/851, or 1027.7/851).

We further checked for the model with a velocity shift in the absorber along our line of sight (with the convolution model "zmshift" in XSPEC). We found that the model with a shift absorber with a different velocity can further reduce ${\Delta}\chi^{2}$=26.0 for two extra d.o.f.   We also found that the shift partial absorber (with the convolution model "partcov" in XSPEC) with a different velocity was not required.  It can only further reduce ${\Delta}\chi^{2}$=3.8 for one extra d.o.f, with a partial covering factor of $f_c=0.56^{+0.18}_{-0.33}$ for xmmobs1 and $f_c=1$ for xmmobs2. However, we cannot exclude the partial covering scenario from the spectral fittings. We also checked other values of turbulent velocity $v_{turb}$ ranging from 100 to 10,000${\rm \,km\,s^{-1}}$. The results are broadly consistent with the present ones. We listed all the fitting results from a median turbulent velocity value of 3,000${\rm \,km\,s^{-1}}$ in Table\,\ref{tb:shiftwarmref}.

Finally, we obtained a best-fit result with a shifting warm absorbed reflection model ($\chi^{2}/d.o.f$=877.5/845). The X-ray spectra with the best-fit model are shown in Figure\,\ref{fig:warmabsp}. We found a reflection component with a power-law index for the incident spectrum of $\Gamma$=1.59$^{+0.03}_{-0.04}$, a high-energy cut-off of E$_{cut}$ =96.8$^{+31.6}_{-23.7}$ keV, the ionization of the reflection disk of $log[\xi_{r}]=3.27^{+0.17}_{-0.17}$,  an inclination angle of $i =21{^{+8}_{-11}}^{\circ}$, and a reflection factor of  {\rm f$_{r}=0.49^{+0.44}_{-0.08}$}. As the high-energy cut-off of E$_{cut}$ cannot be constrained well by the model and since the obtained value is consistent with those found in the {\it Swift}/BAT AGNs sample \citep{Ricci2017}, we further fixed the E$_{cut}$ at 100~keV.

We note that the warm absorber with an H column density and ionization parameter of $n_{\rm H}\sim7.6\times$10$^{20}$ cm$^{-2}$ and  $log\,[\xi_{w}/ {\rm (erg\,cm\,s^{-1})}]\sim$0.8 is required for xmmobs1. In xmmobs2, the H column density and ionization parameter of the warm absorber are found to be nH$\sim3.8\times$10$^{21}$ cm$^{-2}$ and $log\,[\xi_{w}/ {\rm (erg\,cm\,s^{-1})}]\sim$1.9, respectively. Interestingly, an ultra-fast outflowing velocity of $0.193^{+0.014}_{-0.023} c$ is found for xmmobs1, but the outflowing velocity decreases to $\sim 0.021c$ for xmmobs2. This suggests that the warm absorber has dramatic variations in the column density, ionization parameter, and outflowing velocity, though the X-ray continuum flux has only changed by $\sim$10$\%$.

\section{Summary and discussion}\label{sum_dis}

We present an analysis of the X-ray observations of the unprecedented SMBH binary candidate predicted to merge within three years in a nearby galaxy SDSSJ1430+2303, covering a period from 2021 November 23 to 2022 June 4 by {\it Swift}, {\it XMM-Newton}, {\it Chandra,} and {\it NuSTAR}.  Dramatic variability was found in the X-ray light curve spanning $\sim$200 days. No significant X-ray variability on short timescales of a few hours or shorter was found, as suggested by the two {\it XMM-Newton} observations. The detailed joint analysis of the high-quality X-ray spectra taken by {\it XMM-Newton} (50 and 75~ks), {\it NuSTAR} (100~ks), and {\it Chandra} (57.6~ks) revealed notable spectral features including a warm absorber, broad Fe~K emission, and a high-energy cut-off. We finally adopted a warm absorbed relativistic reflection model to describe the broadband X-ray spectra in the 0.2-70~keV band.

Both the obtained photon index ($\rm \Gamma\sim$1.6) and high-energy cut-off ($\rm E_{cut}\sim$100 keV) of the reflection component are similar to those of other X-ray bright AGNs. The reflection factor ($\rm f_{r}\sim$0.5) and ionization ($\rm log\,[\xi_r]\sim$3.2) also agree well with that given by \citet{Pasham2022} from a preliminary analysis of the {\it NICER} and {\it NuSTAR} data. These results strongly indicate an accretion disk origin for the reflection. Moreover, the low inclination angle ($\rm i\sim20^{\circ}$) implies that the accretion disk is approximately viewed face-on. This is consistent with the VLBI observations \citep{An2022b} in which a single compact core is detected, and is also consistent with the type~1 AGN classification of SDSSJ1430+2303, even though it was initially discovered by nuclear MIR outbursts \citep[]{Jiang2021}. 

\begin{figure}[htb]
\centering
\begin{minipage}{0.48\textwidth}
\centering{\includegraphics[angle=0,width=0.99\textwidth]{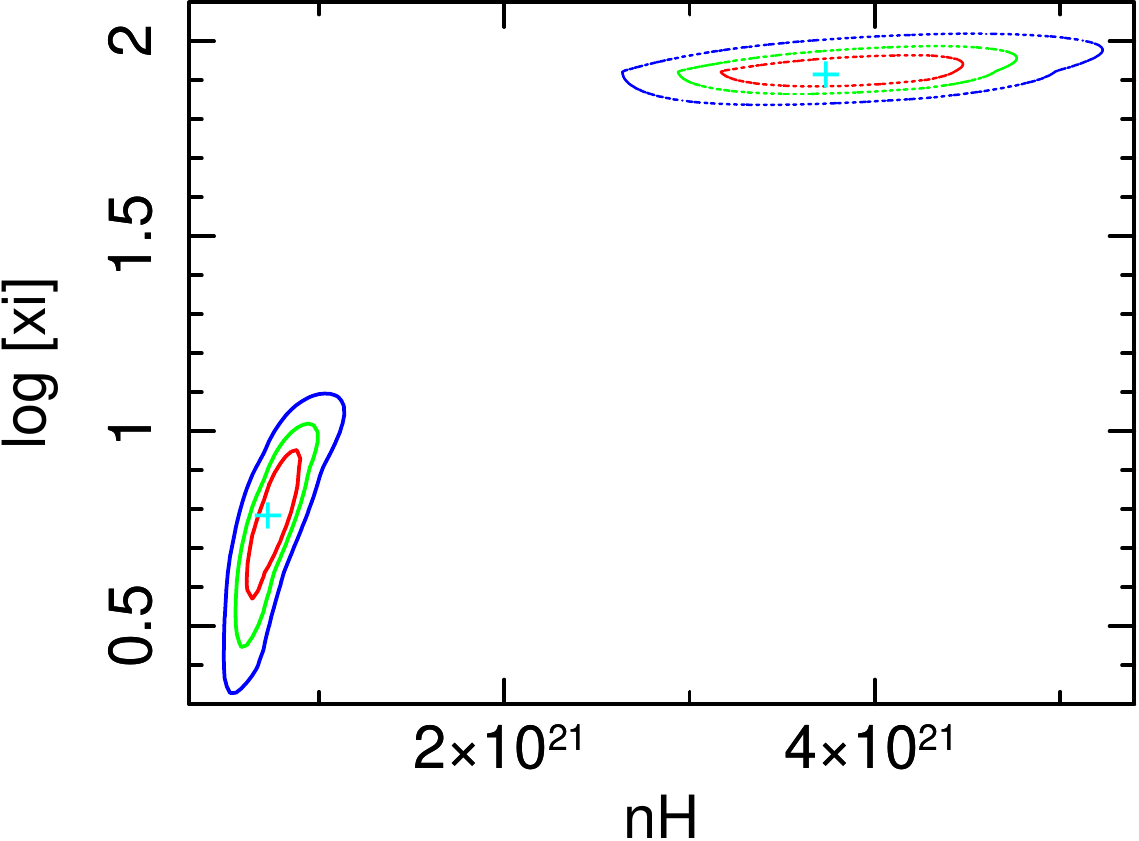}}
\centering{\includegraphics[angle=0,width=0.99\textwidth]{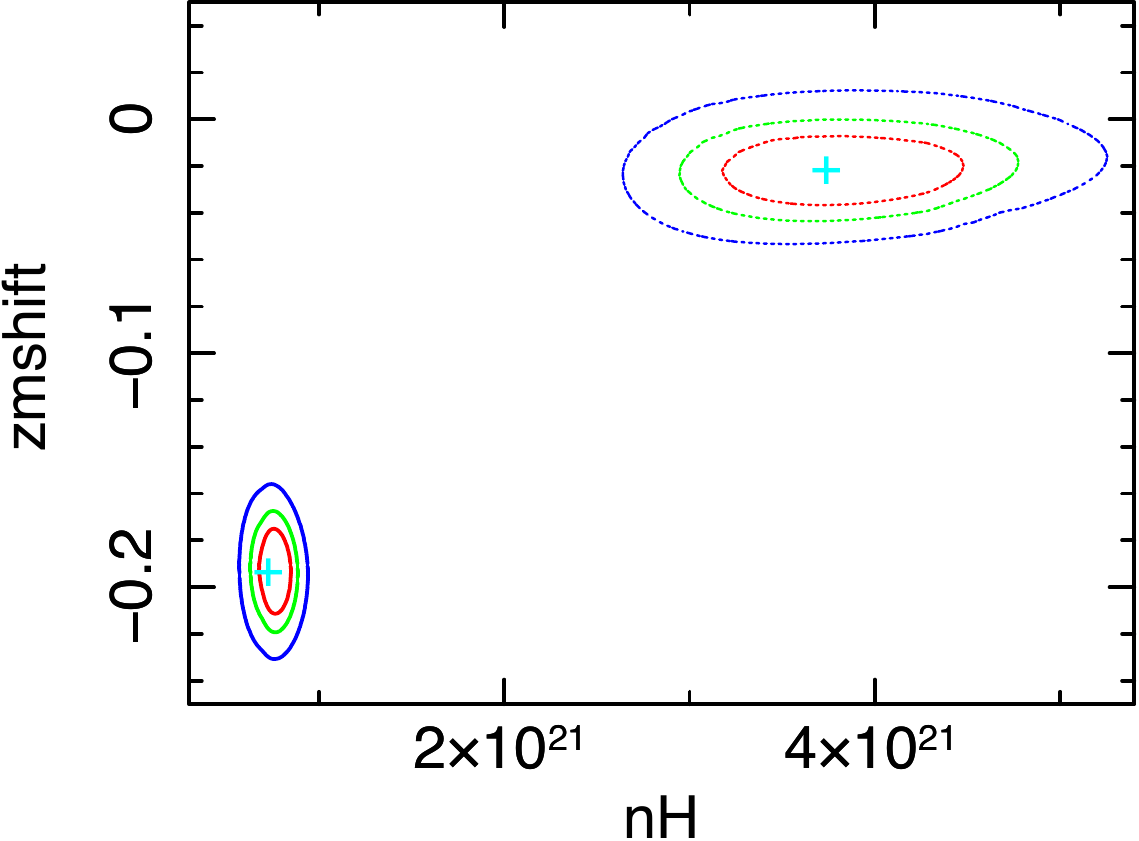}}
\centering{\includegraphics[angle=0,width=0.99\textwidth]{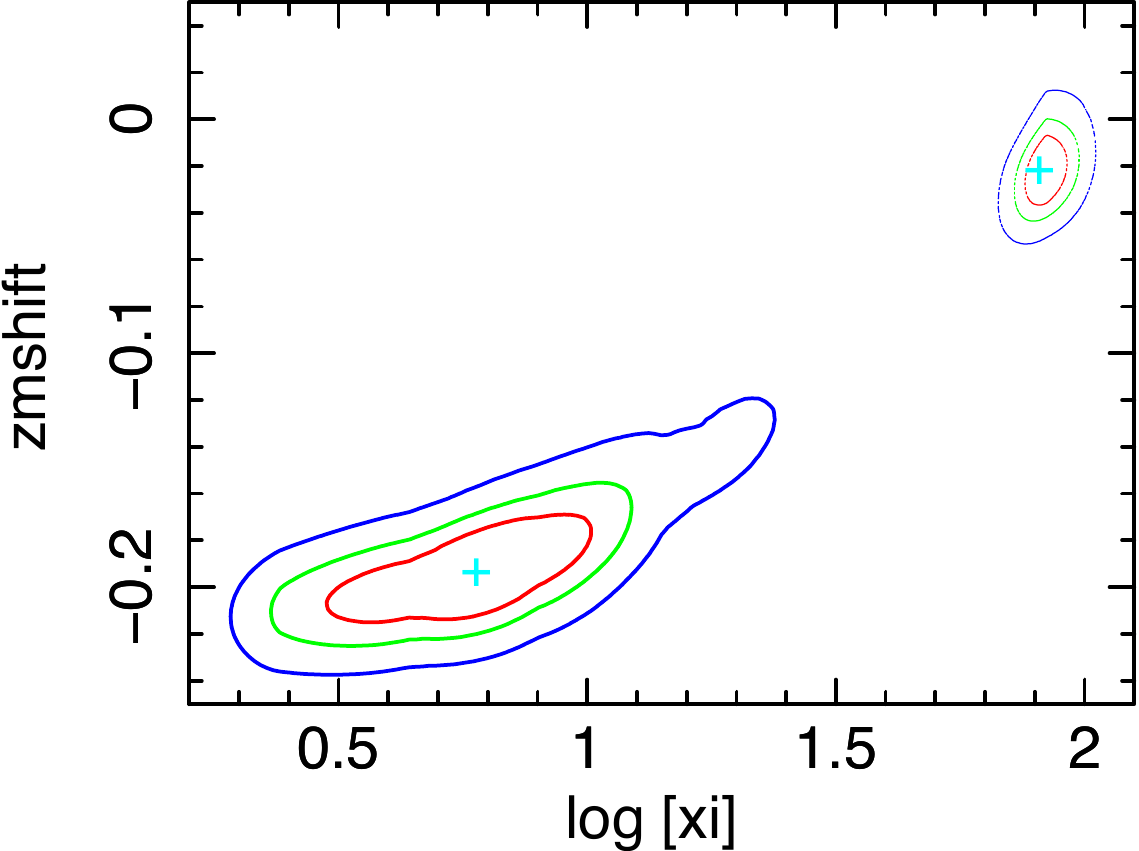}}
\end{minipage}
\caption{Contour plots for the parameters of the warm absorber. The red, green, and blue line represent the $1\sigma$ (68\%), $2\sigma$ (95\%), and $3\sigma$ (99.7\%) confidence levels, respectively. The bottom left is for xmm\_obs1, while the top right is for xmm\_obs2 in each panel.\label{fig:contourwarmabs}}
\end{figure}

For the warm absorber, we found clear variability in the column density, ionization parameter, and outflowing velocity between the two {\it XMM-Newton} observations (see Figure\,\ref{fig:contourwarmabs}) within only $\sim19$ days. In particular, the outflowing velocity has changed from $\sim0.193c$ to $\sim0.021c$, accompanied by a change in the ionization parameter $log\,[\xi_{w}/ {\rm (erg\,cm\,s^{-1})}]$ from 0.8 to 1.92. This is somewhat surprising since the variability in the X-ray continuum flux is marginal (it changed by only $\sim$10$\%$).

The outflowing, even ultra-fast ($v>0.1c$) outflowing warm absorber has been widely found in X-ray luminous AGNs. \citet{Winter2012} detected a warm absorber in about half of a sample of 48 Seyfert galaxies. \citet{Tombesi2013} suggest  that the ultra-fast outflowing absorber is present in about 35$\%$ of nearby Seyfert galaxies.  Such an outflow or ultra-fast outflow is usually interpreted as due to the wind clumps being ejected with different velocities, inclinations, or launching radii from the accretion disk. The physical mechanisms driving the wind could be associated with radiation pressure and/or a magnetic field. Interestingly, the warm absorber in SDSSJ1430+2303 displays a lower ionization parameter at a higher outflowing velocity. Such a phenomenon has rarely been seen alone in previous studies of AGNs \citep[e.g.,][]{Tombesi2013,Laha2014}. A low-ionized-high-velocity outflow accompanying a high-ionized-high-velocity outflow was only detected in a few AGNs, such as IRAS~17020+4544~\citep{Longinotti2015} and PG~1114+445~\citep{Serafinelli2019}.  Even though the possibility of clumpy disk wind in typical AGNs cannot be completely excluded, the phenomenon in SDSSJ1430+2303 is inconsistent with the predictions from radiation-driven or magneto-hydrodynamic-driven winds~\citep[$v_{out} \propto \xi^{0.5}$ or $v_{out} \propto \xi$,][]{King2003,Fukumura2010}. However, the abnormality is not difficult to understand in the SMBHB scenario, that is to say the observed high-velocity outflowing gas could be simply associated with either the orbital motion of the secondary SMBH or the gas that is kicked out during the disk crossing. It would thus be worthwhile to further characterize the temporal variations of the warm absorber.

Lastly, the broad Fe~K$\alpha$ emission is detected at a confidence level of 99.96\% in both {\it Chandra} and {\it XMM-Newton} observations. The equivalent width of the broad Fe~K$\alpha$ emission is found in the range of $\sim$190--320 eV. Including the data from simultaneous {\it NuSTAR} observations, the broad X-ray spectra in the 0.2--70 keV can be self-consistently described by a relativistic reflection model. However, the potential velocity shift or profile changes of the broad Fe~K$\alpha$ emission, which can be considered as the smoking-gun evidence for the SMBH binary scenario, cannot be measured between the two {\it XMM-Newton} observations due to the poor spectral quality. We still encourage further experiments with deeper X-ray observations, particularly at the phase with lower continuum levels. X-rays appear to be the most powerful probe to test the SMBHB interpretation or other proposals \citep{Dotti2022}, especially in  the anticipated final inspiraling stage until the GW memory signal is detected. 

\begin{acknowledgements}\small
We are grateful to the anonymous referee for his/her constructive comments and suggestions which have improved the paper significantly. 
This work is based on observations obtained with {\it XMM-Newton}, an ESA science mission with instruments and contributions directly funded by ESA Member States and NASA, the {\it Chandra} X-ray Observatory, Neil Gehrels {\it Swift} Observatory, the Nuclear Spectroscopic Telescope Array Mission ({\it NuSTAR}) and the Neutron star Interior Composition ExploreR ({\it NICER}). We are grateful to the PI or project scientist of {\it Swift}, {\it XMM-Newton}, {\it Chandra} and {\it NuSTAR} for approving our ToO/DDT requests, as well as the operation and scheduling teams of all involved facilities. 
This work is supported by NSFC (11833007, 12041301, 12073025, 12192221, 12022303, U1731104, U2031106), the B-type Strategic Priority Program of the Chinese Academy of Sciences (Grant No. XDB41000000) and China Manned Spaced Project (CMS-CSST-2021-B11). H.Y. and Z.P. acknowledge supports by the Natural Sciences and Engineering Research Council of Canada and in part by Perimeter Institute for Theoretical Physics. Research at Perimeter Institute is supported in part by the Government of Canada through the Department of Innovation, Science and Economic Development Canada and by the Province of Ontario through the Ministry of Colleges and Universities. Y.A. acknowledge supports by NSFC 12133001 and  Guangdong Basic and Applied Basic Research Foundation 2022A1515012151.  \\

$Facilities$: {\it XMM-Newton}, {\it Chandra}, {\it NuSTAR}, {\it Swift}, {\it NICER}.
$Softwares$: {SAS}, {CIAO}, {HEASOFT}, {XSPEC}.

\end{acknowledgements}




\setcounter{figure}{0} 
\renewcommand{\thefigure}{A\arabic{figure}}
\setcounter{table}{0}
\renewcommand{\thetable}{A\arabic{table}}

\begin{table*}[htb]
\centering
\caption{Power-law fitting result for the spectra in 2-5 keV of two {\it XMM-Newton} and {\it Chandra}, 3-5 keV of {\it NuSTAR}, and 8-10 keV of two {\it XMM-Newton} and {\it NusTAR} observations.}
\label{plfit}
\centering
\begin{tabular}{c c c c c}
\hline\hline
{Observations}  & {$\Gamma$} & {$F_{2-10 keV}$}  & {$L_{2-10 keV}$} & {$\chi^2/d.o.f$} \\
{}  & {} & {10$^{-12}$ erg cm$^{-2}$ s$^{-1}$}  & {10$^{43}$ erg s$^{-1}$} & {} \\
\hline\hline
XMM-obs1        & 1.65$\pm$0.07 & 2.85    & 4.53 & \multirow{4}{*}{373.6/347}  \\
XMM-obs2        & 1.61$\pm$0.07 & 2.62   & 4.14 &   \\ 
Chandra     & 1.65$\pm$0.10 & 2.51   & 3.98 &   \\
NuSTAR      & 1.71$\pm$0.07 & 3.72    & 5.92 &   \\
\hline
\multicolumn{5}{c}{jointly fitting} \\
\hline
       & 1.66$\pm$0.04 &   &   & 376.2/350 \\
\hline
\end{tabular}
\end{table*}

\begin{table*}[htb]
\scriptsize
\caption{Warm absorbed reflection modeling result.}
\label{tb:shiftwarmref}
\centering
\begin{tabular}{c cccc ccccc c}
\hline\hline
{} & \multicolumn{4}{c}{warm absorber (warmabs)}& \multicolumn{5}{c}{reflection (relxill)} &  \\
{Observations}  & {zmshift} & {$n_{\rm H}$} & {$log\,[\xi_{w}]$}  & {\rm $f_{c}$} & {$\Gamma$} & {$log\,[\xi_{r}]$} & {$E_{cut}$}& {\rm $f_{r}$} & {$i_{incl}$} &{$\chi^2/d.o.f$} \\
{}  &{} & {10$^{21}\,cm^{-2}$} & {$erg\,cm\,s^{-1}$}  & {} & {} & {} & {keV}& {} & {$^\circ$} & \\
\hline
\multicolumn{11}{c}{{\it XMM-Newton}, {\it Chandra} and {\it NuSTAR} jointly fitting} \\
\hline
XMM-obs1 & \multirow{2}{*}{$0$} & $1.07^{+2.13}_{-0.32}$  & $1.25^{+0.40}_{-0.22}$  & \multirow{2}{*}{$1$}  & \multirow{2}{*}{$1.56^{+0.03}_{-0.02}$}  & \multirow{2}{*}{$3.31^{+0.05}_{-0.06}$} & \multirow{2}{*}{$79.9^{+21.3}_{-10.2}$} & \multirow{2}{*}{$0.60^{+0.16}_{-0.09}$} & \multirow{2}{*}{$<21$} & \multirow{2}{*}{$903.5/847$} \\
XMM-obs2+C+N &  & {$2.80^{+0.80}_{-0.81}$} & {$1.92^{+0.08}_{-0.06}$} &  &  & & & &  &  \\
\hline
XMM-obs1 & \multirow{2}{*}{$-0.030^{+0.013}_{-0.016}$} & $2.00^{+1.43}_{-1.13}$  & $1.48^{+0.17}_{-0.40}$  & \multirow{2}{*}{$1$}  & \multirow{2}{*}{$1.58^{+0.03}_{-0.03}$}  & \multirow{2}{*}{$3.30^{+0.10}_{-0.13}$} & \multirow{2}{*}{$95.0^{+25.0}_{-22.7}$}  & \multirow{2}{*}{$0.50^{+0.13}_{-0.06}$} & \multirow{2}{*}{$19^{+8}_{-14}$} & \multirow{2}{*}{$893.5/846$}  \\
XMM-obs2+C+N & & {$3.24^{+1.08}_{-0.80}$} & {$1.92^{+0.06}_{-0.12}$} &  &  & & & &  &       \\
\hline
XMM-obs1 & {$-0.193^{+0.014}_{-0.023}$} & $0.76^{+0.28}_{-0.26}$  & $0.79^{+0.25}_{-0.45}$  & \multirow{2}{*}{$1$}  & \multirow{2}{*}{$1.59^{+0.03}_{-0.04}$}  & \multirow{2}{*}{$3.27^{+0.18}_{-0.18}$} & \multirow{2}{*}{$96.8^{+31.6}_{-23.7}$}  & \multirow{2}{*}{$0.49^{+0.44}_{-0.08}$} & \multirow{2}{*}{$25^{+8}_{-11}$} & \multirow{2}{*}{$877.5/845$}  \\
XMM-obs2+C+N & {$-0.022^{+0.017}_{-0.017}$} & {$3.81^{+0.92}_{-1.03}$} & {$1.92^{+0.05}_{-0.08}$} &  &  & & & &  &       \\
\hline
XMM-obs1 & {$-0.193^{+0.028}_{-0.024}$} & $0.76^{+0.26}_{-0.44}$ & $0.79^{+0.25}_{-0.45}$  & \multirow{2}{*}{$1$} & \multirow{2}{*}{$1.59^{+0.03}_{-0.02}$}  & \multirow{2}{*}{$3.27^{+0.19}_{-0.17}$} & \multirow{2}{*}{$100$}  & \multirow{2}{*}{$0.49^{+0.44}_{-0.08}$} & \multirow{2}{*}{$25^{+8}_{-12}$} & \multirow{2}{*}{$877.5/846$}  \\
XMM-obs2+C+N & {$-0.021^{+0.016}_{-0.017}$} & {$3.77^{+1.09}_{-0.86}$} & {$1.92^{+0.06}_{-0.08}$} &  &  & & & &  &       \\
\hline
XMM-obs1 & {$-0.205^{+0.032}_{-0.032}$} & $1.57^{+1.08}_{-0.86}$  & $0.75^{+0.23}_{-0.24}$  & \multirow{2}{*}{$0.59^{+0.40}_{-0.16}$}  & \multirow{2}{*}{$1.61^{+0.02}_{-0.05}$}  & \multirow{2}{*}{$3.22^{+0.24}_{-0.21}$} & \multirow{2}{*}{$100$}  & \multirow{2}{*}{$0.43^{+0.32}_{-0.10}$} & \multirow{2}{*}{$<28$} & \multirow{2}{*}{$874.8/845$}  \\
XMM-obs2+C+N & {$-0.023^{+0.022}_{-0.020}$} & {$8.04^{+5.64}_{-4.72}$} & {$1.90^{+0.11}_{-0.07}$} & & & & & & & \\
\hline
XMM-obs1 & {$-0.200^{+0.032}_{-0.032}$} & $1.63^{+1.38}_{-0.83}$ & $0.74^{+0.18}_{-0.23}$ & {$0.56^{+0.18}_{-0.33}$} & \multirow{2}{*}{$1.60^{+0.03}_{-0.05}$} & \multirow{2}{*}{$3.24^{+0.24}_{-0.19}$} & \multirow{2}{*}{$100$} & \multirow{2}{*}{$0.46^{+0.45}_{-0.08}$} & \multirow{2}{*}{$<30$} & \multirow{2}{*}{$873.7/845$} \\
XMM-obs2+C+N & {$-0.023^{+0.018}_{-0.018}$} & {$3.94^{+1.24}_{-1.06}$} & {$1.92^{+0.06}_{-0.09}$} & $1$ &  & & & & & \\
\hline
\end{tabular}
\end{table*}

\begin{figure*}[htb]
\centering
\begin{minipage}{0.45\textwidth}
\centering{\includegraphics[angle=0,width=1.0\textwidth]{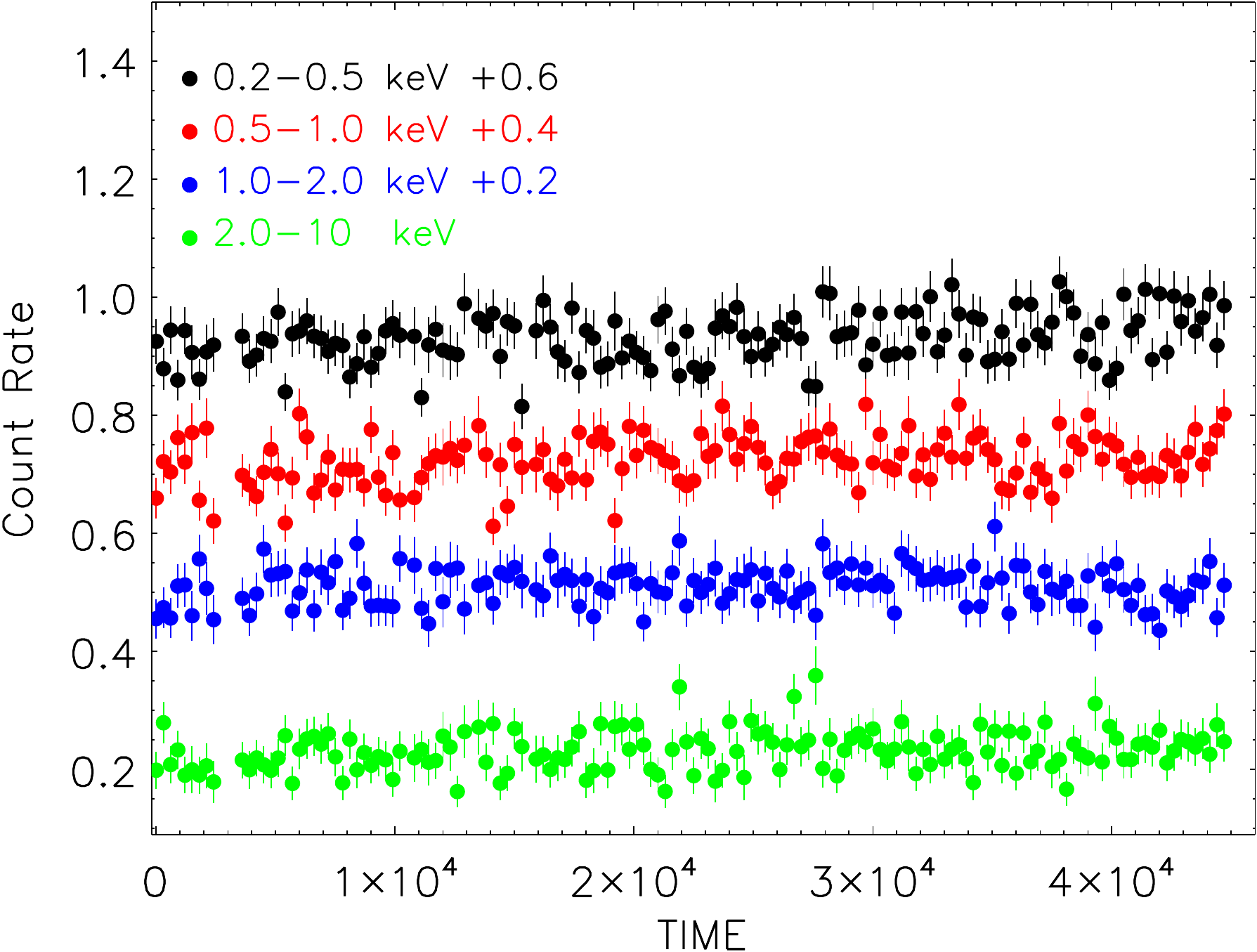}}
\end{minipage}
\begin{minipage}{0.45\textwidth}
\centering{\includegraphics[angle=0,width=1.0\textwidth]{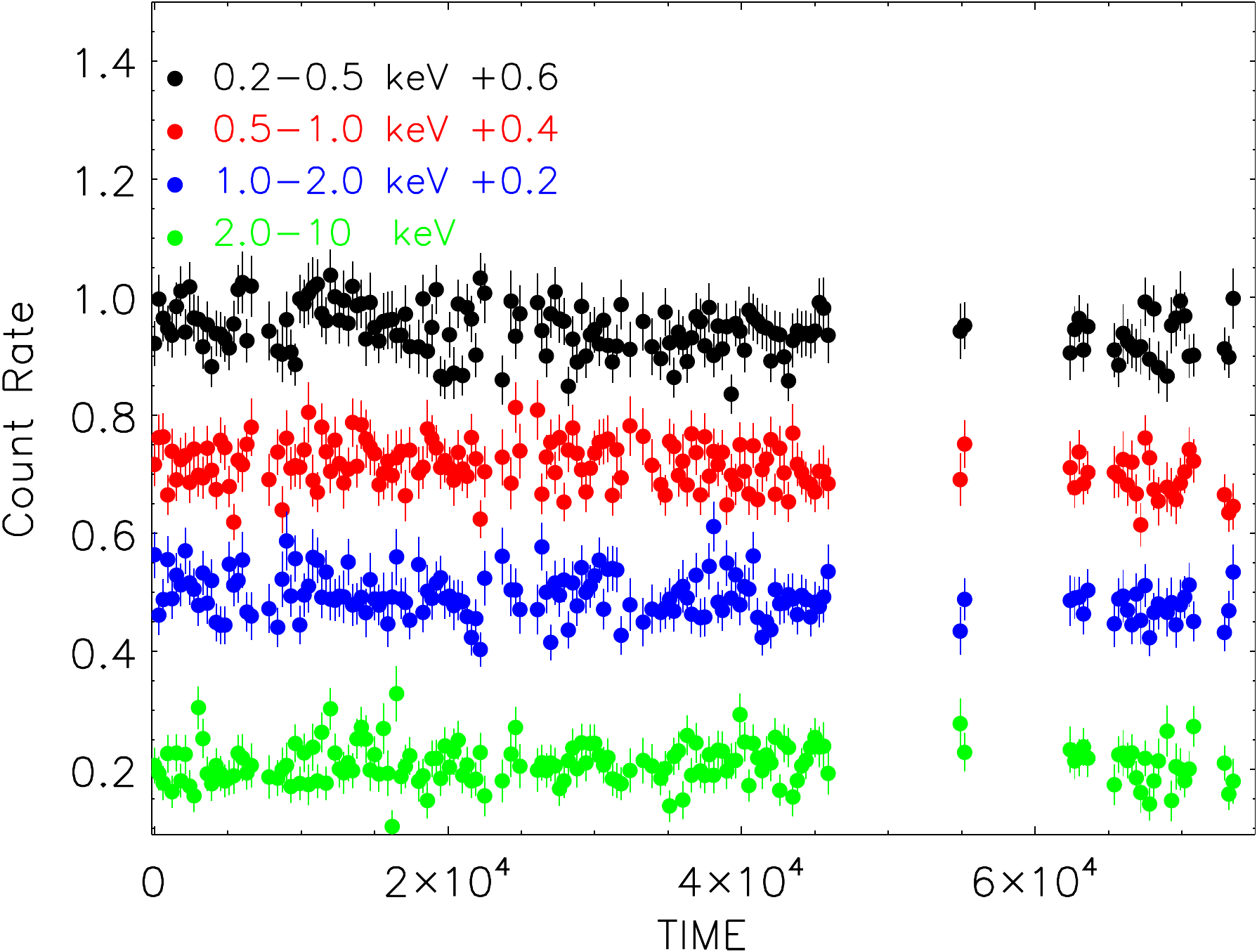}}
\end{minipage}
\caption{X-ray light curves of the two {\it XMM-Newton}/pn observations with a time bin size of 300 seconds in each 0.2-0.5, 0.5-1, 1-2, and 2-10 keV band. Left panel: The observation on 2021 December 31 (xmmobs1). Right panel: The observation on 2022 January 19 (xmmobs2).
\label{xlc_o1o2}}
\end{figure*}


\begin{thebibliography}{}

\bibitem[An et al.(2022a)]{An2022a} An, T., Zhang, X., Wang, A., et al.\ 2022, The Astronomer's Telegram, 15267
\bibitem[An et al.(2022b)]{An2022b} An, T., Zhang, Y., Wang, A., et al.\ 2022, \aap, 663, A139. 
\bibitem[Arzoumanian et al.(2014)]{Arzoumanian2014} Arzoumanian, Z., Gendreau, K.~C., Baker, C.~L., et al.\ 2014, \procspie, 9144, 914420. 
\bibitem[Begelman et al.(1980)]{Begelman1980} Begelman, M.~C., Blandford, R.~D., \& Rees, M.~J.\ 1980, \nat, 287, 307
\bibitem[Boroson \& Lauer(2009)]{Boroson2009} Boroson, T.~A. \& Lauer, T.~R.\ 2009, \nat, 458, 53
\bibitem[Charisi et al.(2016)]{Charisi2016} Charisi, M., Bartos, I., Haiman, Z., et al.\ 2016, \mnras, 463, 2145
\bibitem[Chen et al.(2020)]{Chen2020} Chen, Y.-C., Liu, X., Liao, W.-T., et al.\ 2020, \mnras, 499, 2245
\bibitem[Comerford et al.(2015)]{Comerford2015} Comerford, J.~M., Pooley, D., Barrows, R.~S., et al.\ 2015, \apj, 806, 219
\bibitem[D'Orazio et al.(2015)]{D'Orazio2015} D'Orazio, D.~J., Haiman, Z., \& Schiminovich, D.\ 2015, \nat, 525, 351
\bibitem[Dotti et al. (2022)]{Dotti2022} Dotti, M., Bonetti, M., Rigamonti, F., et al. arXiv: 2205.06275
\bibitem[Eracleous et al.(2012)]{Eracleous2012} Eracleous, M., Boroson, T.~A., Halpern, J.~P., et al.\ 2012, \apjs, 201, 23
\bibitem[Fukumura et al.(2010)]{Fukumura2010} Fukumura, K., Kazanas, D., Contopoulos, I., et al.\ 2010, \apj, 715, 636 
\bibitem[Gallimore \& Beswick(2004)]{Gallimore2004} Gallimore, J.~F. \& Beswick, R.\ 2004, \aj, 127, 239
\bibitem[Graham et al.(2015)]{Graham2015} Graham, M.~J., Djorgovski, S.~G., Stern, D., et al.\ 2015, \nat, 518, 74
\bibitem[Graham et al.(2015)]{Graham2015b} Graham, M.~J., Djorgovski, S.~G., Stern, D., et al.\ 2015, \mnras, 453, 1562
\bibitem[G{\"u}ltekin \& Miller(2012)]{Gultekin2012} G{\"u}ltekin, K. \& Miller, J.~M.\ 2012, \apj, 761, 90
\bibitem[Garc{\'\i}a et al.(2014)]{Garca2014} Garc{\'\i}a, J., Dauser, T., Lohfink, A., et al.\ 2014, \apj, 782, 76. 
\bibitem[Hayasaki(2009)]{Hayasaki2009} Hayasaki, K.\ 2009, \pasj, 61, 65
\bibitem[Haehnelt(1994)]{Haehnelt1994} Haehnelt, M.~G.\ 1994, \mnras, 269, 199
\bibitem[HI4PI Collaboration et al.(2016)]{HI4PI2016} HI4PI Collaboration, Ben Bekhti, N., Flöer, L., et al.\ 2016, \aap, 594, A116. 
\bibitem[Ivanov et al.(1999)]{Ivanov1999} Ivanov, P.~B., Papaloizou, J.~C.~B., \& Polnarev, A.~G.\ 1999, \mnras, 307, 79
\bibitem[Jaffe \& Backer(2003)]{Jaffe2003} Jaffe, A.~H. \& Backer, D.~C.\ 2003, \apj, 583, 616
\bibitem[Jiang et al.(2021)]{Jiang2021} Jiang, N., Wang, T., Dou, L., et al.\ 2021, \apjs, 252, 32
\bibitem[Jiang et al.(2022)]{Jiang2022} Jiang, N., Yang, H., Wang, T., et al.\ 2022, arXiv:2201.11633
\bibitem[Kallman \& Bautista(2001)]{Kallman2001} Kallman, T. \& Bautista, M.\ 2001, \apjs, 133, 221. doi:10.1086/319184
\bibitem[King(2003)]{King2003} King, A.\ 2003, \apjl, 596, L27.
\bibitem[Kormendy \& Ho(2013)]{KH13} Kormendy, J. \& Ho, L.~C.\ 2013, \araa, 51, 511
\bibitem[Laha et al.(2014)]{Laha2014} Laha, S., Guainazzi, M., Dewangan, G.~C., et al.\ 2014, \mnras, 441, 2613. 
\bibitem[Lehto \& Valtonen(1996)]{Lehto1996} Lehto, H.~J. \& Valtonen, M.~J.\ 1996, \apj, 460, 207
\bibitem[Liu et al.(2010)]{Liu2010} Liu, X., Greene, J.~E., Shen, Y., et al.\ 2010, \apjl, 715, L30
\bibitem[Liu et al.(2016)]{Liu2016} Liu, T., Gezari, S., Burgett, W., et al.\ 2016, \apj, 833, 6
\bibitem[Longinotti et al.(2015)]{Longinotti2015} Longinotti, A.~L., Krongold, Y., Guainazzi, M., et al.\ 2015, \apjl, 813, L39.
\bibitem[MacFadyen \& Milosavljevi{\'c}(2008)]{MacFadyen2008} MacFadyen, A.~I. \& Milosavljevi{\'c}, M.\ 2008, \apj, 672, 83
\bibitem[Nandra et al.(2007)]{Nandra2007} Nandra, K., O'Neill, P.~M., George, I.~M., et al.\ 2007, \mnras, 382, 194.
\bibitem[Noble et al.(2012)]{Noble2012} Noble, S.~C., Mundim, B.~C., Nakano, H., et al.\ 2012, \apj, 755, 51
\bibitem[Pasham et al.(2022)]{Pasham2022} Pasham, D., Fabian, A., Walton, D., et al.\ 2022, The Astronomer's Telegram, 15225
\bibitem[Protassov et al.(2002)]{Protassov2002} Protassov, R., van Dyk, D.~A., Connors, A., et al.\ 2002, \apj, 571, 545.
\bibitem[Runnoe et al.(2017)]{Runnoe2017} Runnoe, J.~C., Eracleous, M., Pennell, A., et al.\ 2017, \mnras, 468, 1683
\bibitem[Ricci et al.(2017)]{Ricci2017} Ricci, C., Trakhtenbrot, B., Koss, M.~J., et al.\ 2017, \apjs, 233, 17. 
\bibitem[Serafinelli et al.(2019)]{Serafinelli2019} Serafinelli, R., Tombesi, F., Vagnetti, F., et al.\ 2019, \aap, 627, A121.
\bibitem[Shen et al.(2013)]{Shen2013} Shen, Y., Liu, X., Loeb, A., et al.\ 2013, \apj, 775, 49
\bibitem[Sillanpaa et al.(1988)]{Sillanpaa1988} Sillanpaa, A., Haarala, S., Valtonen, M.~J., et al.\ 1988, \apj, 325, 628
\bibitem[Winter et al.(2012)]{Winter2012} Winter, L.~M., Veilleux, S., McKernan, B., et al.\ 2012, \apj, 745, 107. 
\bibitem[Thorne \& Braginskii(1976)]{Thorne1976} Thorne, K.~S. \& Braginskii, V.~B.\ 1976, \apjl, 204, L1
\bibitem[Tombesi et al.(2013)]{Tombesi2013} Tombesi, F., Cappi, M., Reeves, J.~N., et al.\ 2013, \mnras, 430, 1102. 
\bibitem[Yan et al.(2015)]{Yan2015} Yan, C.-S., Lu, Y., Dai, X., et al.\ 2015, \apj, 809, 117
\bibitem[Zheng et al.(2016)]{Zheng2016} Zheng, Z.-Y., Butler, N.~R., Shen, Y., et al.\ 2016, \apj, 827, 56
\bibitem[Zhou et al.(2004)]{Zhou2004} Zhou, H., Wang, T., Zhang, X., et al.\ 2004, \apjl, 604, L33.

\end{thebibliography}
\end{document}